\documentclass[twocolumn,runningheads]{svjour2}
\smartqed  
\usepackage{graphicx}

\newcommand{\etal}{et al. }
\newcommand{\fun}{{\rm cm^{-2}\, s^{-1}}}
\journalname{Astrophysics and Space Science}
\begin{document}

\title{Unidentified EGRET Sources and the Extragalactic Gamma-Ray 
Background \thanks{This work was supported by the Kavli Institute for 
Cosmological Physics through the grant NSF PHY-0114422 and by 
DOE grant DE-FG0291-ER40606 at the University of Chicago.}
}

\author{Vasiliki Pavlidou         \and
        Jennifer M. Siegal-Gaskins \and
        Carolyn Brown \and \\
        Brian D. Fields  \and 
        Angela V. Olinto 
}

\institute{V. Pavlidou, J. M. Siegal-Gaskins, 
C. Brown, A.V. Olinto \at
The University of Chicago, Chicago, IL 60637 
          \and
           B.D. Fields \at
University of Illinois at Urbana-Champaign, Urbana, IL 61801
}

\date{Received: date / Accepted: date}

\maketitle

\begin{abstract}

The large majority of EGRET point sources remain to this day 
without an identified 
low-energy counterpart. Whatever the nature of the EGRET unidentified 
sources, faint unresolved objects of the same class must have a 
contribution to the diffuse gamma-ray background: if most unidentified 
objects are extragalactic, faint unresolved sources of the same class 
contribute to the background, as a distinct extragalactic population;  
on the other hand, if most unidentified sources are Galactic, their 
counterparts in external galaxies will contribute to the unresolved 
emission from these systems. Understanding this component 
of the gamma-ray background, along with other guaranteed contributions 
from known sources, is essential in any attempt to use  
gamma-ray observations to constrain
exotic high-energy physics. Here, we follow an empirical approach 
to estimate whether a potential contribution of unidentified 
sources to the extragalactic gamma-ray background is likely to be 
important, and we find that it is. Additionally, we 
comment on how the anticipated GLAST measurement of the diffuse 
gamma-ray background will change, depending on the nature of the 
majority of these sources.

\keywords{Gamma rays: observations \and 
Gamma-ray sources: astronomical \and Radiation sources: unidentified}
\PACS{95.85.Pw \and 98.70.Rz \and 98.70.-f }
\end{abstract}

\section{Introduction}
\label{intro}

The EGRET telescope aboard the {\it Compton Gamma-Ray Observatory}
detected, during its nine years of operation, not only
271 point sources \cite{3rdcat} but also
diffuse emission \cite{gal,ext}. 
Based on the origin of the diffuse photons, we can classify this
emission as Galactic if it is produced within the Milky Way, or
as extragalactic if it originates from larger, cosmological distances.  
The two types of
emission can be identified, to a certain extent, 
 from their spatial distribution:
Galactic emission is expected to be enhanced near the Galactic plane,
while extragalactic emission is expected to be largely
isotropic -- however, the Galactic emission does not quickly fall off
to zero as we move away from the plane. 
Disentangling the Galactic and extragalactic components requires
modeling the Galactic emission through cosmic ray propagation models 
(e.g. \cite{gal,strong}) and subtracting it from the all-sky diffuse signal. 
The uncertainties involved in this process are nontrivial.
Residuals from inadequate 
or imperfect treatment of the dominant Galactic component 
can severely contaminate the determination of the extragalactic 
emission \cite{dar,keshet,strong}.

Diffuse emission can also be characterized, depending on the process
by which it is produced, as  truly diffuse or as unresolved point
source emission.  Truly diffuse is, for example,  the 
emission resulting from cosmic ray interaction with interstellar
matter in the Milky Way\footnote{Note however that enhanced emission 
close to cosmic-ray sources could in principle be resolved 
as point source emission and would thus count as ``unresolved 
point source emission'' in this classification scheme.}, 
as well as the emission from particles
accelerated in shocks at the outskirts of 
cosmological large-scale structures. 
In contrast, unresolved point source
emission is the emission produced by a collection of faint, unresolved
point sources such as blazars, perceived as diffuse due to
limitations in telescope sensitivity. 

Although many sources have been suggested to be the origin of 
the diffuse emission, there are some {\em guaranteed} contributions. 
Any known class of gamma-ray sources with some
already identified members 
must have at least some contribution to diffuse emission. This would
originate from the collective emission from fainter members,
unresolved by EGRET. 
Prime examples of such classes are blazars, which
 may be a dominant component of the extragalactic
diffuse emission, e.g. \cite{SS96,mc99,mp00,nt06}, 
as well as normal galaxies, e.g. \cite{l78,PF02}, 
or pulsars (for the case of the Galactic diffuse emission), e.g. 
\cite{hs81,petal,zc}.

It is also possible that unresolved sources of the same class as 
unidentified EGRET sources have some appreciable contribution to 
the extragalactic background. Although their nature remains unknown,
it is reasonable to believe that there is a large
number of fainter, unresolved objects of the same class, making some
contribution to the diffuse emission. In addition, unidentified sources 
are the most numerous group of gamma-ray sources. If they represent
yet unidentified members of some known class of gamma-ray emitters
(e.g. blazars), then excluding them from any calculation of 
the contribution of the parent class to the diffuse background
would lead to a significantly underestimated result, 
due to an incorrect normalization of the bright-end of the
gamma-ray luminosity function. 
If they represent an
unknown class of gamma-ray emitters (which is likely -- see e.g.
\cite{reimaip}), then the contribution of their
unresolved counterparts to the diffuse emission would significantly
limit the diffuse flux left to be attributed to known classes
and to truly diffuse emission. 

Similarly,
investigating the diffuse emission from unresolved unidentified
objects is important regardless of the location of these objects
(whether they are Galactic or extragalactic). If they are 
extragalactic,  then unresolved objects of the same class 
contribute to the diffuse extragalactic gamma-ray
background. Alternatively, if they are all Galactic,  
objects of the same class in other unresolved 
galaxies enhance the contribution of their hosts to 
the  gamma-ray background\footnote{Additionally, unresolved 
sources of the same class in our own galaxy contribute to the 
diffuse emission  of the Milky Way.}. 

Hence, some contribution of unresolved unidentified sources to the 
extragalactic diffuse
background is certain. It is therefore clear that until we either (a) 
resolve the issue of the nature of unidentified sources or (b) derive
some strong constraint indicating that a possible contribution of such
unresolved objects would indeed be minor, we cannot hope to be
confident in our understanding of the origin of the extragalactic 
gamma-ray background. 
However, predictions for the level of their collective contribution  
involve important uncertainties: due to
lack of identification of low-energy counterparts, we have no estimates
of distance, and therefore no estimates of the gamma-ray 
luminosities of these sources. For this reason,
very few constraints can be placed on their cosmic distribution 
and evolution. 

In this work, we approach the problem from a purely empirical
point of view. Instead of attempting to {\em predict} the level of a
diffuse component due to unresolved objects of the same class as
unidentified EGRET sources, we try to assess whether there are any
quantitative indications that this component is, in fact, minor. 
Under the assumption that the majority of the
unidentified EGRET sources can be treated as members of a single 
class of gamma-ray emitters, and for the case that this class 
consists of {\em extragalactic objects}, we try to answer
the following two questions:
(1) is it likely that unresolved objects of the same class could have
    a significant contribution to the extragalactic gamma-ray
    background at least in some energy range, and 
(2) how would the collective spectrum of their emission compare to the
    measured spectrum of the extragalactic gamma-ray background
    deduced from EGRET observations. 
The observational input constraining our calculations will be the
number distribution of unidentified sources with respect to flux, and
the spectral index of each source. 
We also examine how we expect GLAST observations to change our knowledge of
the nature of unidentified sources, based on the insight gained from
our analysis.  

This paper is structured as follows. In \S \ref{sec:form} we summarize
the formalism used to derive the extragalactic gamma-ray background
component due to unresolved unidentified sources under our set of
assumptions. In \S \ref{sec:res} we describe our results and their
implications.  Prospects for the GLAST era are discussed in \S
\ref{sec:GLAST}. Finally, we conclude and discuss our findings in 
\S \ref{sec:disc}.

\section{Formalism}

In this section, we describe how we can use the flux distribution of 
unidentified sources to extract information about a possible 
contribution of unresolved sources of the same class to the
extragalactic diffuse gamma-ray background. 
The assumptions we will make to proceed are that (a) 
the unidentified sources can indeed
be viewed as a class of objects, so that constructing a flux
distribution is meaningful\footnote{In reality, even if most of the
unidentified sources do belong to the same class, there is always
going to be some contamination of different class objects; however, it
is conceivable that such contamination is small.}; and (b) close
to the EGRET flux limit, the flux distribution does not evolve 
drastically so that an extrapolation of the measured flux distribution to lower
fluxes is representative of its behavior in the low-flux regime. This
assumption is less likely to hold as the limiting flux to which we are
extrapolating becomes lower: the flux distribution
{\em will} eventually  exhibit a
break due to cosmological effects and/or luminosity evolution.

 The question we will seek to answer is
how far in the low-flux regime our extrapolation must continue 
before we get a significant contribution of unresolved unidentified 
sources to the gamma-ray background. 
If the answer is ``not very far'', then we might expect
that the actual flux distribution of the
unresolved sources does indeed  
resemble our assumed form and that unresolved, unidentified sources
make up a considerable fraction of the extragalactic diffuse
background. On the other hand, if we need to extrapolate the flux
distribution down to fluxes very low compared to the resolved flux
range, then it is quite unlikely that our extrapolation is valid
throughout the flux regime we are using it. In such a case it is 
doubtful that the actual flux distribution  
of the unresolved unidentified sources is
indeed such that unidentified sources make up a significant portion of
the extragalactic diffuse background. 
\label{sec:form}
\begin{figure}
\centering
\medskip
  \includegraphics[width=0.38\textwidth]{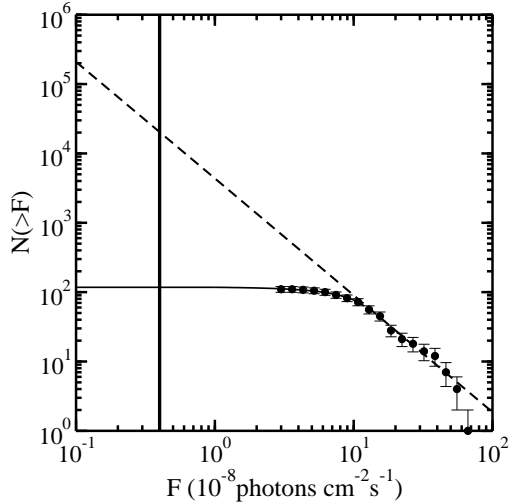}
\caption{Flux distribution of unidentified sources. The error bars
represent $1\sigma$ Poisson errors. Dashed line: power-law fit in the
range $13\times 10^{-8} \, \fun \leq F \leq 55 \times 10^{-8}\,
\fun$. Solid line: parabolic fit in the
range $3 \times 10^{-8} \,\fun \leq F \leq 13 \times 10^{-8} \,
\fun$. 
Thick solid line: $F_{\rm min}$.}
\label{fig:1}       
\end{figure}

\subsection{The cumulative flux distribution}

In our calculation, we will only include
 in the ``unidentified object'' class
those 3rd EGRET catalog sources which  still remain
without a suggested low-energy counterpart 
candidate\footnote{Sample compiled and
 actively maintained  
by C. Brown and available online
at \\ http://home.uchicago.edu/$\sim$carolynb/unidentified\_sources. 
It is important to note that suggested associations with lower-energy
counterparts included in this compilation are simply results reported
in recent publications; no effort was made to evaluate the
significance and validity of these counterparts by a single uniform standard.}.
If any sources originally included in the 3rd EGRET catalog listing of
unidentified sources have been omitted when they should in fact have
been included in our object sample, the error that this omission
incurs is consistently toward the side of {\em underestimating} the
importance of the contribution of unidentified sources to the diffuse
background. The opposite extreme would be  
if we had chosen to include all sources marked as
unidentified in the 3rd EGRET catalog. 
Including 
{\em more} sources in the resolved sample than we did, 
would significantly enhance the bright-end of the flux distribution; 
this would also imply more faint-end, 
unresolved, unidentified sources, which would support a stronger unidentified
source component in diffuse radiation.
In a future version of this work, we plan to assess quantitatively 
the uncertainties introduced by our
choice of the ``unidentified sources'' sample 
by examining to what extent our conclusions may change when different 
sets of EGRET  sources, satisfying different criteria, are used.

The cumulative flux distribution, $N(>F)$ (number of
sources with flux above $F$ versus $F$) of these sources is plotted in
Fig. \ref{fig:1}. 
Here, $F$ is the mean\footnote{Note that we have not tried to address
strong variability of the sources.} (P1234) photon flux in energies above 
$100 \, {\rm MeV}$ 
quoted in the 3rd EGRET catalog \cite{3rdcat}. The dashed line
is the power-law fit to the data in the 
flux interval between $13 \times 10^{-8}
\, \fun$ and $55 \times 10^{-8} \, \fun$, 
$N_b(>F) = CF_8^{-\kappa}\,,$
with $F_{8} = F/ (10^{-8} \, \fun)$, $\ln C=8.32 \pm 0.36 $ 
and $\kappa = 1.67 \pm 0.11$. In the ``faint source'' end of the 
cumulative flux distribution, the data necessarily deviate from the
power law fit due to the finite sensitivity of the telescope. 
However,  because of differences in sky coverage, the 
angular dependence of the diffuse background, and
source variability, there is no sharp cutoff in the fluxes of resolved
objects. Instead, the cumulative flux distribution gradually flattens.
The thin solid line shows a polynomial 
fit to the data in the faint source end (flux interval
between $3 \times 10^{-8} \, \fun$ and $13 \times 10^{-8} \, \fun$), 
$N_f(>F) = A_0- A_1 F_{8} - A_2 F_{8}^2\,,$
 with $A_0 =  119.5 $, $A_1 = 1.837$ and $A_2 = 0.24214$. 
The subscripts $b$ and $f$ in $N(>F)$ refer to the ``bright'' and
``faint'' end of the cumulative flux distribution respectively. 

If an extrapolation of the power law fit to low
fluxes is representative of the number of existing sources 
in the faint source end, then the {\em differential} flux distribution 
(number of objects with fluxes between $F$ and $F+dF$)
of existing sources (both resolved and unresolved) of the
``unidentified class'' in units of $({10^{-8} \, \fun})^{-1}$ is 
$\left|dN/dF\right|_b = \kappa C F_8^{-\kappa-1}$.
In the faint end of source fluxes, the number of {\em
unresolved} sources with fluxes between $F$ and $F+dF$ 
is the number
of existing sources minus the number of resolved sources, or 
\begin{equation}\label{eq:unresolved}
\left|\frac{dN}{dF}\right|_u dF = 
dF_8
\left[\kappa C F_8^{-\kappa-1} - A_1 - 2 A_2 F_8\right]\,.
\end{equation}

\subsection{The spectral index distribution}\label{isid}

The second important observational input in our calculation 
is the distribution of spectral indices of unidentified
objects. In our analysis we will adopt the assumption that the
spectral index distribution of unresolved unidentified sources is the
same as that of the resolved unidentified sources. The latter 
can be deduced from measurements of the spectral index $\alpha$
for each of the
resolved unidentified sources. 

\begin{figure}
\bigskip
\bigskip
\centering
  \includegraphics[width=0.34\textwidth]{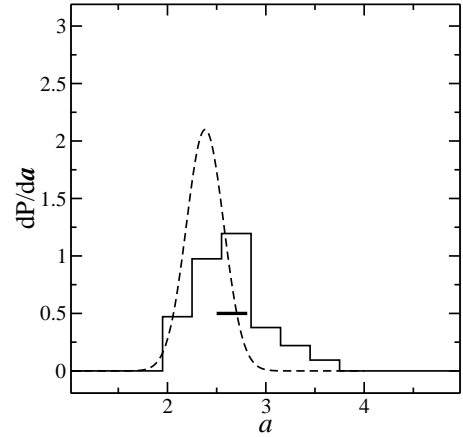}
\caption{Spectral index distribution of unidentified sources. Solid
line: histogram of the EGRET data. Dashed line: maximum-likelihood
gaussian. Thick solid line: typical uncertainty of individual spectral index
determination using EGRET data. }
\label{fig:15}      
\end{figure}

Figure \ref{fig:15} shows 
a histogram of the spectral indices of the
resolved unidentified sources (solid line). Note that the typical 
measurement uncertainty for any single spectral index (thick solid line)
is  comparable with the spread of the
distribution, so that the spread of a simple binning of spectral 
indices might not
in fact give us information about the underlying distribution of the
spectral indices of the sources, but rather be representative
of the uncertainty of each single measurement.  
This problem is not unique to unidentified sources. It is also a
problem  in measuring the spectral index distribution of blazars,
where the  usually derived concavity \cite{SS96} of the collective unresolved
blazar spectrum may simply be the result of overestimating the spread in
the spectral  index distribution \cite{VP06}.

Following an analysis similar to that of \cite{VP06}
for the case of blazars, we assume that the
intrinsic spectral index distribution of unidentified sources can be
approximated by a gaussian.
We then use a  maximum-likelihood  analysis which takes 
into account the individual errors of 
measurement of $\alpha$ for each source\footnote{We do 
this by introducing the true spectral 
indices of the sources as nuisance parameters and by marginalizing 
over them.} to estimate the parameters
of the distribution, obtaining
a mean of $\alpha_0 = 2.38 \pm 0.03$ and a standard deviation of 
$\sigma_\alpha = 0.19 \pm
0.03$.
The errors quoted
are $1\sigma$ uncertainties of each parameter with the other fixed at
its maximum-likelihood value. The maximum-likelihood distribution is
plotted with the dashed line in Fig. \ref{fig:15}, and  
is narrower and displaced to lower spectral indices with respect to
the histogram (a result reflecting the fact that sources with higher
values of $\alpha$ also tend to have a larger error of measurement in
$\alpha$). 
However, due to the large systematics associated with the
observational determinations of the extragalactic gamma-ray background
spectrum at high energies, our conclusions are not very sensitive to the
small displacement of the peak.

\subsection{Contribution to the extragalactic gamma-ray background}

The extragalactic gamma-ray background is described by the
differential photon intensity $I_E$ (photons per unit
area-time-energy-solid angle).
Each  unresolved source of flux $F$ has a contribution 
$I_{E,1}(F)$ to the diffuse emission  which is given by 
\begin{equation}\label{single}
I_{E,1}(F) = (\alpha-1)\frac{F}{4\pi E_0} 
\left(\frac{E}{E_0}\right)^{-\alpha}\,, 
\end{equation}
where $\alpha$ is the spectral index of the source, and $E_0 = 100
{\rm \, MeV}$ is the lowest photon energy included in the measurement
of $F$. The $1/4\pi$ normalization factor comes from assuming an
isotropic distribution of sources, the collective emission of which is
uniformly distributed over the celestial sphere. 
Using Eq. (\ref{eq:unresolved}) for the distribution of fluxes, 
the maximum-likelihood gaussian of \S \ref{isid} for the spectral 
index distribution, and integrating over 
the desired flux interval and over all possible spectral indices, 
we can then calculate 
the collective diffuse emission due to unresolved sources of the
``unidentified class'' with fluxes between $F_{\rm min}$ and $F_{\rm max}$.
We will take $F_{\rm max} = 13 \times 10^{-8} \, \rm \fun$
(the lower end of the ``bright source'' flux interval, where we assume
that all sources with fluxes above this limit have been resolved by
EGRET). $F_{\rm min}$ (the low-flux limit of the flux-function extrapolation) 
is the free parameter in our calculation. 

\section{Results}\label{sec:res}
Integrating Eq. (\ref{single}) we can now calculate the contribution of
unresolved unidentified sources to the diffuse background as a
function of $F_{\rm min}$. Note that the overall level of the 
associated diffuse
emission only depends on where the extrapolated power law breaks. 
The thick solid line in Fig. \ref{fig:1} shows
the value of this cutoff so that the contribution of unidentified
sources does not overtake the original EGRET estimate of the 
extragalactic gamma-ray background \cite{ext} in the
range 100-300 MeV plus statistical error\footnote{This is a 
conservative choice, since all more recent estimates of the
extragalactic diffuse background give 
an EGRB level lower than the original estimate of \cite{ext}. If
we had instead adopted a lower EGRB intensity, the thick solid line 
of Fig. \ref{fig:1} would move to higher fluxes, and this would make an
important unidentified source component of the EGRB even more likely.}.
We would only need extrapolate the cumulative flux distribution for
slightly more than an order of magnitude below
the lower limit of the resolved flux range to have
unresolved unidentified sources comprise most of the extragalactic
diffuse background, at least at low energies. This is not an extreme 
extrapolation, and therefore a significant 
contribution by the ``unidentified'' class to the diffuse background
is likely.  

\begin{figure}
\centering
  \includegraphics[width=0.42\textwidth]{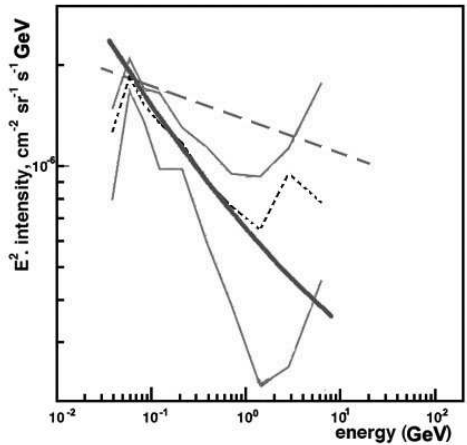}
\caption{Dashed line: Sreekumar \etal (1998) determination of the
EGRB. Dotted line: Strong \etal (2004) determination of the EGRB (best
guess). Solid lines: Strong \etal (2004) determination of the EGRB
(systematics-based limits). Thick solid line: collective spectrum of
unresolved unidentified sources (this work). }
\label{fig:2}       
\end{figure}

Figure \ref{fig:2} shows the cumulative emission spectrum of unresolved
unidentified sources, overplotted
with the spectrum of the extragalactic diffuse emission derived
from EGRET observations. The dashed line shows a single--power-law fit
to the Sreekumar et al. (1998) determination of the extragalactic
gamma-ray background (EGRB) \cite{ext}.  
The dotted line 
is the more recent redetermination of the gamma-ray background by  
Strong et al (2004) \cite{strong}, in which they used 
their more detailed model of the Milky Way diffuse emission to 
subtract the Galactic component from the EGRET diffuse sky map. 
The solid lines are  the systematic uncertainties in the EGRB 
determination of \cite{strong}, 
entering through their model of the Galaxy. Our 
calculation of the spectrum of the unresolved unidentified source
component is shown with the thick solid line.
At low energies, where the systematics are low, the unidentified
component spectrum is in excellent agreement with the
EGRB observational spectrum of \cite{strong}. At higher energies, 
where the systematics are
large, the unidentified component spectrum is largely within
systematics except at very high energies. 
If unidentified sources are indeed a dominant contribution at
relatively low energies, then this result may be perceived as a
tantalizing hint that at the
highest energies of the EGRET range a new type of contribution,
(e.g. from annihilating dark matter) may start to be important at a few 
tens of GeV. 
 
\section{Prospects for the GLAST era}\label{sec:GLAST}

The launch of GLAST in 2007 will provide us with significant new
insight about the nature of unidentified sources and their possible
contribution to the extragalactic diffuse background. 
The ideal solution to the unidentified source puzzle would be, of
course, the direct positional association of all unidentified sources
with undisputed low-energy counterparts.  This would then allow us to
build more confident models for the unresolved members of these
classes of objects. However, such an outcome is unlikely, as 
the large number of possible counterparts and the large number of
sources which we expect GLAST will be able to resolve make
multi-wavelength campaigns for every single source impractical. 

However, there is another definitive test that GLAST will be able to
perform, which does not require confident identification of each
source to provide information about the likely nature
of unidentified sources as a population. With the increased 
flux sensitivity of GLAST, many more objects of the same class will be
resolved. 
If these objects are mostly extragalactic, there will be an
associated decrease of the extragalactic gamma-ray background from its
EGRET levels, equal 
to the all-sky-averaged intensity of the newly resolved objects. 
The flux sensitivity of GLAST is expected to be about 50 times
better than that of EGRET\footnote
{http://www-glast.stanford.edu/mission.html}. 
Therefore GLAST will be able to probe the
flux distribution of unidentified sources down to fluxes close to
$F_{\rm min}$ and definitively test our empirical estimate. 

If, on the other hand,  these objects are mostly Galactic, then there
will be an associated reduction of the Milky Way diffuse emission
rather than of the isotropic background.  
For a discussion on a possible contribution of a large number of 
Galactic point sources to the Galactic diffuse emission and the 
role of such sources in
explaining at least in part the origin of the GeV excess see
\cite{strong2}.

\section{Discussion}\label{sec:disc}

In this work, we have used a purely empirical model to explore
 the possibility that unresolved
gamma-ray sources of the same class as unidentified EGRET sources have
an appreciable contribution to the extragalactic gamma-ray
background. We have argued that some unidentified source contribution 
to the gamma-ray background is guaranteed. We have additionally found
that 
(1) if most unidentified sources are assumed to be extragalactic, 
    we would only need the observed cumulative flux distribution of
    unidentified sources to  extend without a break for a little more
    that one order of magnitude toward lower fluxes in order to have a  very
   significant contribution to the gamma-ray background -- at least at
   the lower part of the EGRET energy range; and 
(2) the spectrum of the cumulative emission of such unresolved sources
    would be very consistent with the observational
    determination of \cite{strong} of the gamma-ray background 
    from EGRET data, within systematics. 

We have learned that any model of the
extragalactic gamma-ray background would be incomplete without some
treatment of the unidentified source contribution. The results of our
empirical model therefore motivate us 
to pursue specific population models for the unidentified sources. 
Although such models involve  a more restrictive set of
assumptions and increased uncertainties, they can provide more 
concrete predictions for the luminosity function of unresolved
objects. Additionally, if we were to assume that the majority of
unidentified sources are indeed members of a single, extragalactic  
class of gamma-ray emitters, and that unresolved members of
this class do indeed contribute most of the extragalactic diffuse
emission, then we can use simple evolution models to place limits on
the redshifts of unidentified sources. 
We will pursue such models and calculations in an upcoming publication.  

In this work we have tried, where possible, to make 
assumptions, which, if anything, {\em underestimate} the possible
contribution of unresolved unidentified sources to the extragalactic
diffuse background. However, some of our necessary working assumptions
have the potential to overestimate the unidentified class
contribution. 

First of all, we have assumed that the majority of the sources belong
in a single class. It is conceivable that instead, the resolved
unidentified sources are a collection of members of several known and
unknown classes of gamma-ray emitters. In this case, it is
still likely that the summed contribution of unresolved members of all
parent classes to the diffuse background is significant. However, the
construction of a single cumulative flux distribution from all sources 
and its extrapolation to lower fluxes is no longer an indicative test
for the importance of such a contribution. 


Second, we have assumed that the majority of the sources we have used 
are extragalactic. It is not at all certain that this is indeed the
case. The unidentified sources from the 3rd EGRET catalog
\cite{3rdcat} exhibit a 
strong concentration along the Galactic plane, and hence a significant
fraction of the unidentified sources in the 3rd EGRET catalog are 
most likely Galactic. In the sample we have used, 
this feature is less pronounced
(mainly because recent suggestions of possible low-energy counterparts
refer mostly to Galactic plane objects). As a result, the possibility 
that {\em most} of the sources we have used in our
calculations are extragalactic cannot be excluded on isotropy
grounds. However, a {\em mostly} Galactic population originating in
the disk and bulge is also consistent with constraints from the
gamma-ray emission from nearby galaxies \cite{S06j}. In this
case the total luminosity of {\em resolved} unidentified sources is
already a significant fraction of the total diffuse luminosity of
normal galaxies. Consequently the luminosity of each individual normal 
galaxy is significantly enhanced and the unidentified sources have a
contribution to the extragalactic background through their hosts. 
Ultimately, the question of whether most unidentified sources are
Galactic or extragalactic  will be decided by GLAST. 

\begin{acknowledgements}
We are indebted to S. Gabici, I. Grenier, M. Longair, 
T. Prodanovi\'{c}, O. Reimer,
A. Strong, K. Tassis, and T. Venters for comments and 
discussions related to different aspects
of this work.
\end{acknowledgements}



\end{document}